\documentstyle{mn}

\newif\ifAMStwofonts

 \ifoldfss
  \ifCUPmtlplainloaded \else
    \NewTextAlphabet{textbfit} {cmbxti10} {}
    \NewTextAlphabet{textbfss} {cmssbx10} {}
    \NewMathAlphabet{mathbfit} {cmbxti10} {} 
    \NewMathAlphabet{mathbfss} {cmssbx10} {} 
  \fi
  \ifAMStwofonts
    \ifCUPmtlplainloaded \else
      \NewSymbolFont{upmath} {eurm10}
      \NewSymbolFont{AMSa} {msam10}
      \NewMathSymbol{\upi}     {0}{upmath}{19}
      \NewMathSymbol{\umu}     {0}{upmath}{16}
      \NewMathSymbol{\upartial}{0}{upmath}{40}
      \NewMathSymbol{\leqslant}{3}{AMSa}{36}
      \NewMathSymbol{\geqslant}{3}{AMSa}{3E}

    \fi
  \fi
\fi 

\ifnfssone
  \newmathalphabet{\mathit}
  \addtoversion{normal}{\mathit}{cmr}{m}{it}
  \addtoversion{bold}{\mathit}{cmr}{bx}{it}
  \newmathalphabet{\mathbfit} 
  \addtoversion{normal}{\mathbfit}{cmr}{bx}{it}
  \addtoversion{bold}{\mathbfit}{cmr}{bx}{it}
  \newmathalphabet{\mathbfss} 
  \addtoversion{normal}{\mathbfss}{cmss}{bx}{n}
  \addtoversion{bold}{\mathbfss}{cmss}{bx}{n}
  \ifAMStwofonts
    \ifCUPmtlplainloaded \else
      \UseAMStwoboldmath
      \makeatletter
      \new@mathgroup\upmath@group
      \define@mathgroup\mv@normal\upmath@group{eur}{m}{n}
      \define@mathgroup\mv@bold\upmath@group{eur}{b}{n}
      \edef\UPM{\hexnumber\upmath@group}
      \new@mathgroup\amsa@group
      \define@mathgroup\mv@normal\amsa@group{msa}{m}{n}
      \define@mathgroup\mv@bold\amsa@group{msa}{m}{n}
      \edef\AMSa{\hexnumber\amsa@group}
      \makeatother
      \mathchardef\upi="0\UPM19
      \mathchardef\umu="0\UPM16
      \mathchardef\upartial="0\UPM40
      \mathchardef\leqslant="3\AMSa36
      \mathchardef\geqslant="3\AMSa3E
    \fi
  \fi
\fi 

\ifnfsstwo
  \DeclareMathAlphabet{\mathbfit}{OT1}{cmr}{bx}{it}
  \SetMathAlphabet\mathbfit{bold}{OT1}{cmr}{bx}{it}
  \DeclareMathAlphabet{\mathbfss}{OT1}{cmss}{bx}{n}
  \SetMathAlphabet\mathbfss{bold}{OT1}{cmss}{bx}{n}
  \ifAMStwofonts
    \ifCUPmtlplainloaded \else
      \DeclareSymbolFont{UPM}{U}{eur}{m}{n}
      \SetSymbolFont{UPM}{bold}{U}{eur}{b}{n}
      \DeclareSymbolFont{AMSa}{U}{msa}{m}{n}
      \DeclareMathSymbol{\upi}{0}{UPM}{"19}
      \DeclareMathSymbol{\umu}{0}{UPM}{"16}
      \DeclareMathSymbol{\upartial}{0}{UPM}{"40}
      \DeclareMathSymbol{\leqslant}{3}{AMSa}{"36}
      \DeclareMathSymbol{\geqslant}{3}{AMSa}{"3E}
    \fi
  \fi
\fi 

\ifCUPmtlplainloaded \else
  \ifAMStwofonts \else 
    \def\upi{\pi}
    \def\umu{\mu}
    \def\upartial{\partial}
  \fi
\fi

\title[A search for brown-dwarf secondaries in CVs]{A search for brown-dwarf like secondaries in cataclysmic
variables
\thanks{
Based on observations
obtained at the European Southern Observatory, ESO proposal 65.H-0409(A).}}
\author[R.E.\ Mennickent and M.P. Diaz]
       {R. E. Mennickent$^1$\thanks{E-mail: rmennick@stars.cfm.udec.cl} and M. P. Diaz$^2$\\
$^1$Dpto. de F\'{\i}sica, Facultad de Ciencias F\'{\i}sicas y Matem\'aticas,
Universidad de Concepci\'on, Casilla 160-C, Concepci\'on, Chile\\
$^2$Instituto  Astron\^omico e Geof\'{\i}sico,
Universidade de S\~ao Paulo, Brazil}
\date{Accepted .
      Received ;
      in original form 2002 February 1}

\pagerange{\pageref{firstpage}--\pageref{lastpage}}
\pubyear{1994}

\begin{document}

\maketitle

\label{firstpage}

\begin{abstract}
  We present VTL/ISAAC infrared spectroscopy of a sample of
short orbital period cataclysmic variables which are candidates
for harboring substellar companions.
We have detected the \hbox{K\,{\sc i}} and \hbox{Na\,{\sc i}} absorption
lines of the companion star in \hbox{VY Aqr}. The overall
spectral distribution in this system is best fit with 
a M9.5 type dwarf spectra,
implying a distance of $100 \pm 10$ pc.
\hbox{VY Aqr} seems to fall far from the
theoretical distribution of secondary star temperatures around the orbital
period minimum.
Fitting of the IR spectral energy distribution (SED) was performed by
comparing the observed spectrum with late-type templates.
The application of such a spectral fitting procedure
suggests that the continuum shape in the 1.1-2.5 $\mu$m spectral region
in short orbital period cataclysmic variables
may be an useful indicator of the companion spectral type.
The SED fitting for \hbox{RZ Leo} and \hbox{CU
Vel} suggests M5 type dwarf companions, and distances of 340 $\pm$ 110 and
150 $\pm$ 50 pc, respectively. These systems may be placed in the upper
evolution branch for short period cataclysmic variables.
\end{abstract}

\begin{keywords}
Stars: individual: \hbox{VY Aqr}, \hbox{V485 Cen},
\hbox{RZ Leo}, \hbox{WZ Sge}, \hbox{CU Vel}, \hbox{HV
Vir},  \hbox{1RXSJ105010.3}, Stars: binaries: close,
Stars: Cataclysmic Variables, Dwarf novae, fundamental parameters, evolution,
Stars: binaries: general, Stars: low-mass
\end{keywords}

\section{Introduction}
Cataclysmic variable stars (CVs) are semi detached binaries
consisting of an accreting white dwarf and a
red dwarf donor transferring matter to the
compact object via the inner Lagrangian point.
The orbiting gas interacts with itself dissipating energy by viscous forces,
forming a luminous accretion disk around the white dwarf. The spectral
energy distribution of CVs in X-rays and ultraviolet is dominated by
white dwarf and inner disk emission, whereas the accretion disk contribution is
dominant
in the optical and eventually in the near infrared. In some cases the emission
from the accretion disk in the IR is approximated by a power-law (Ciardi et al.
1998). However, in
general we expect a much more complex emission from low-luminosity disks
which may contain
extended optically thin regions. On the other hand, the secondary star might 
contribute significantly in
the infrared. The determination of the secondary mass in CVs is key to
understanding the secular evolution of these objects. Current theories state
that the process of mass transfer becomes linked with the loss of orbital
momentum, so the binary period becomes shorter while 
the hydrogen secondary becomes less and
less massive, eventually being eroded by the process, resulting in a kind
of brown dwarf star when the orbital period approaches 80 minutes (e.g.\,
Howell et al. 2001). An alternative
scenario considers that most CVs may not yet have had time to evolve
to their theoretical minimum orbital periods.
In this case, for initial configurations with  
intermediate secondary masses, thermal-timescale mass transfer
may occur. ``If some nuclear evolution has already occurred, this phase
can shrink the binary drastically and strip the hydrogen-rich envelope
from the donor, ultimately producing an ultrashort-period
system with a low-mass, hydrogen poor and probably degenerate
secondary" (from King \& Schenker 2002).    
From the above it is evident that relevant
observations for probing current theories of CV evolution should focus on the
determination of the physical parameters of the secondary star for systems
below the orbital period gap.
At present, four methods have been used to search for
undermassive secondary stars in
cataclysmic variable stars:
(1) analysis of the spectral energy distribution  using
multi-wave-band observations through the ultra-violet,
optical, and infrared spectral regions
(Ciardi et al. 1998; Mason 2001), (2)
looking  for signatures of the secondary star
(Steeghs et al.\, 2001, Littlefair et al. 2000, Dhillon \& Marsh 1995), (3)
``weighting" the secondary star in systems where the superhump and
orbital periods are known (the primary mass is usually assumed,
see for instance Patterson 2001), and (4)
by  spectroscopic diagnostic of the stellar masses
in systems where the white dwarf is revealed
by their optical absorption wings
(Mennickent et al. 2001). Concerning the first two methods, one should
mention that the determination of basic properties of
secondaries in CVs by comparison of their spectra with field stars calibrations
is intrinsically uncertain. These results are prone to illumination and
heating of the companion photosphere. In addition,
the absorption spectrum may be affected by the filling of some lines with
emission components.

While the IR spectra of CVs has been measured and modeled in the past years
there are only a few spectrophotometric observations of low-luminosity systems
in J,H and K band. In this paper we describe the spectra of 7 CVs with
orbital period below
the period gap. These candidates for systems at
late evolutionary stages 
were selected for observation
using the following criteria:
i. their short orbital period and/or ii. a low mass transfer rate, inferred from
their high-amplitude dwarf-nova outbursts with long recurrence time.
In the next section we describe the IR spectroscopic observations while in
Section 3 a description of each spectra is given. A brief discussion of the
observational results is made in Section 4. A few conclusions and
perspectives are outlined in Section 5.

\section{Observations and data reduction}

The infrared spectroscopic observations reported in this paper were
obtained at ESO with VLT-Antu using ISAAC spectrograph in service mode. Data
were taken under photometric atmospheric conditions.  
An observing log is given in Table 1.
Spectra in the J, H and K bands were obtained. 
Sufficient overlapping was assured for
composing a single spectra ranging from 1.09 to 2.57 microns
with a FWHM resolution of 12 (J) to 27 (K) angstroms.
The data were reduced using IRAF\footnote{IRAF is distributed by the
National Optical Astronomy Observatories, which are operated by the
Association of Universities for Research in Astronomy, Inc., under
cooperative agreement with the National Science Foundation} by first
applying the combined dark and flatfield  images as supplied by the ESO
service mode operation group. Median sky frames were then combined for
each object and spectral window. This process made use of jittered images
where the object is located at different positions along the slit.
Wavelength calibration was achieved by measuring the location of
atmospheric OH emission lines (Rousselot et. al.\,2000) in the  sky
background. The flux calibration was performed using observations of the
A0 standard HD216009, made with the same instrumental setup (but with a
wider slit) by the ESO operation team as part of the service mode program.
Finally, the telluric absorption features were corrected with the aid of
the absorption template constructed by dividing the spectrum of HD216009
with a low order polynomial fit, excluding a few stellar absorption
lines.  The IRAF task ``telluric" was
used to find the best scale and shift factors which, when applied to the
normalized telluric template, provided a reasonable correction of the
telluric absorptions in HD216009 and science exposures. This procedure
worked well, except for the regions between 1.35-1.44 microns  and
1.80-1.94 microns, characterized by heavy telluric absorption. These
regions, corresponding to the ends of the J, H and K spectra, were excluded
from the analysis and are not shown in this paper.
Synthetic J, H, and K photometry of our calibrated standard star
observations were compared to broadband photometry by Carter \& Meadows
(1995) showing differences below 0.08 magnitudes. Slit losses for our
objects  were estimated by assuming a Gaussian seeing profile centered in
the slit aperture. For that we used the seeing value included in the
report of the observing block and neglected any factor due to
telescope guiding.
These corrections were included in the J, H and K
magnitudes and in the final spectrum as well. We are confident that this
method worked well, since our spectrophotometric magnitudes compare well
with previously published photometry. In addition, we obtain a good 
match in the overlapping region between
J,H and K spectra.

\subsection{Infrared spectral fitting}

An attempt to quantify the properties of the IR spectra was made
by employing a numerical fitting procedure.
The spectral energy distribution (SED) in the IR was tentatively
parameterized by adding the contribution from a late-type template spectrum
and power-law component. Although the emission of the disk should 
differ substantially from a power-law in the IR, we introduced this component
as a first  approximation to the accretion disc continuum.

While the power law is smooth in our wavelength domain and
basically affects the slope of the IR continuum, the detailed shape of
the synthetic continuum in the J, H and K is strongly dependent on the
stellar template contribution.
A nonlinear least squares fitting procedure
 was calculated using the following equation:\\

$ S(\lambda) = a \times T(\lambda) + b \times \lambda^{c}$ \hfill(1)\\

\noindent
where S is the observed spectrum, T the red dwarf template spectrum,
$\lambda$ the wavelength in microns
and $\it{a}$,$\it{b}$,$\it{c}$ parameters to be
found. The parameters $a$, $b$ and $c$ were adjusted to minimize the reduced Chi-square
between the observed spectrum and a model fit. 
Of course, $\it{a}$ and $\it{b}$ are constrained to the positive domain.
The data fit in this way were selected carefully avoiding emission
lines and deep telluric bands.
A sequence of template spectral types between M1 and L7 at
2 subtypes step was taken from Leggett et al. (2001). The 
resolution of the
template spectra (typically between 20$-$35 \AA) was matched by convolving our data with a Gaussian with the appropiate
FWHM.
Due to the low velocity of the templates,
it was not necessary to account for rotational broadening 
when fitting the spectrum.

\section{Results}

The infrared spectra for all program stars
are shown in Fig.\,1.
The synthetic magnitudes
are given in Table 2 and the
emission line parameters in Table 3. Table 4
summarizes the results of the spectral energy distribution
fitting, giving the parameters for the best fitting function
as defined in  Eq.\,1.

\subsection{VY Aqr}

VY Aqr is a cataclysmic variable showing one of the largest
outburst amplitude among dwarf novae (Downes et al. 2001).
Accordingly to current models for dwarf nova outburst (Osaki 1996), this is
consistent with a very low mass transfer rate system.
Spectroscopic studies in the optical region have revealed the
orbital period (0\fd06309(4) Thorstensen 1997), but not  the stellar
masses, likely  due to the biasing nature of the emission line radial
velocities  (e.g. Augusteijn 1994). J-band spectroscopy by Littlefair et
al.\, (2000) revealed spectral features of the secondary star but too weak
to make an estimate of the spectral type. According to the current CV
evolution scenario (e.g. Howell et al 2001), the orbital period and the
low mass transfer rate of VY Aqr should suggest a system beyond the
orbital period minimum, probably containing  an undermassive secondary
star. This is also supported by the difference between observed superhump
period and the orbital period (Patterson 2001). 
In this section we present direct
evidence for this view, based on the unambiguous detection of  the
secondary star in the spectrum of VY Aqr.

We measured synthetic magnitudes J = 15.17, H = 14.66 and K = 14.14
with estimated errors of 0.10 mag, indicating that VY Aqr
was observed in quiescence. 
Our J and K magnitudes compares well
with the 15.24 $\pm$ 0.15 and 14.42 $\pm$ 0.13 values given by
Sproats et al. (1996).
The infrared spectrum of \hbox{VY Aqr} shows
Bracket and Paschen emission lines.
Fig.\,2  clearly shows the \hbox{K\,{\sc i}} doublets at
1.169-1.177 and 1.244-1.253 and the \hbox{Na\,{\sc i}} line at 1.141,
which are signatures
of a cool secondary star, confirming previous indications found by
Littlefair et al. (2000).

From the visual inspection of the spectral features detected in VY Aqr,
we conclude that the secondary in VY Aqr is of spectral type later than M6
because of the following features: 1- the depth of \hbox{K\,{\sc i}}  doublets
increases for later types. Even disregarding veiling, their depth suggests
a type cooler than M5. 2- The depth of water band at 1.32 microns is too
shallow in dwarf spectra earlier than M6. 3- Strong \hbox{Al\,{\sc i}} lines at
1.313 and 1.32 $\mu$m, usually visible in dwarfs hotter than M6, are not
seen in the object spectrum.  On the other hand, a secondary spectrum
earlier than L5 is suggested by: 1- the strength of \hbox{Na\,{\sc i}} 1.141
relative to the \hbox{K\,{\sc i}} doublet at 1.169, 1.177 and 2- the depth of the
water band.

When fitting the spectral energy distribution (Fig\, 3), we observed
that spectral types earlier than M7 fail to reproduce the depth of
\hbox{K\,{\sc i}} lines in the J band and the continuum in the K band.
On the other hand, spectral
types later than L3  do not fit well the H and K band continuum shape.
These cool types present a  well defined CO bandhead at 2.29 $\mu$m which
is not seen in our data. Our fit with a M9.5 type secondary is slightly
better than that for M7 and L3 type templates, giving a $\chi^{2}$
parameter about 15\% lower. If such a spectral contribution is in fact
due to the
emission of the secondary star in the system one may estimate its
temperature.
Using the effective temperatures for L type dwarfs derived
by Leggett et al. (2001) using structural models,
we find $T_{2}$  =
2300 $\pm$ 100 $K$ for the secondary star in \hbox{VY Aqr}.
We note that our result is in apparent disagreement with
the result of 1400 $\pm$ 50 $K$ obtained by Mason (2001)
by modeling the (non-simultaneous) UV-optical-infrared
spectra. However, our result is still below the 2600 $K$ upper limit
she found by fitting her photometric dataset.
Also our result is
consistent with the upper limit suggested by Littlefair et al. (2000).
The best fit with a M9.5 companion is shown for
illustration in Fig.\,3. Representative fits using types between
M7 and L3 indicate
that the secondary star may contribute with 55\% to
45\% of the flux at 2.17 $\mu$m, depending on the spectral type. Later types
yield better fits for smaller flux fractions. Using the distance values for
our templates from M7 to L2 (LHS3003 and LHS429 by van Altena et al. 1995
and Kelu-1 from Dahn et al. 2000)  and the flux fractions derived from
the spectral fitting we were able to derive a distance estimate for the
system between 80 and 120 pc,
with a most likely value of 100 $\pm$ 10 pc.
The spectroscopic parallax given above is in
agreement with the distance of 110 pc found by Augusteijn (1994) using the
average absolute magnitude value for dwarf novae in outburst.

\subsection{V485 Cen} This is a dwarf nova with an ultra-short
orbital period of 59 minutes (Augusteijn et al. 1996). These authors found
strong evidence for a significant  contribution from the secondary to the
spectrum at wavelengths longer than 5900 Angstroms, which was
confirmed by Munari \& Zwitter (1998), based on the observed color
progression. According to
Augusteijn et al. (1996), the most likely explanation for
the extremely short orbital period, it is a non-fully
degenerate secondary star with a low, but finite,
hydrogen content.
Considering our M-L spectral type templates,
the best fit was reached with
a simple power continuum with index -1.7.

\subsection{RZ Leo} The long cycle length and large amplitude
outburst relate this star to \hbox{WZ Sge}, however the orbital period
(0.07651 d, Mennickent \& Tappert 2001) is longer and the spectral
distribution redder (e.g. Mennickent et al. 1999). This fact, along with
the rather normal mass ratio estimated from the superhump period (Ishioka
et al. 2001) point to a normal secondary for this object.

Our data seems to confirm this view. Fig.\,4 shows that the best SED model
for \hbox{RZ Leo} is reached with a M5 type spectrum and almost zero
contribution from the power law. M3 and M7 subtype
templates failed to match the shape of the spectrum, giving, in
both cases, a $\chi^{2}$ three times larger than the M5 template. We do not
see the absorption lines of the secondary in the J-band spectrum probably
due to our poor S/N in this range. Using $M_{J}$ for M dwarfs from
Leggett et al. (2000),
we  estimate  a distance of 340 $\pm$ 110  pc. Our $J$ and $K$ magnitudes,
viz.\, 16.57 and 15.45, are similar to that given by Sproats et al.
(1996), namely 16.56 $\pm$ 0.19 and 15.55 $\pm$ 0.17.

\subsection{WZ Sge} The spectral type of the secondary in this system has been
constrained to be later than M7.5 (Littlefair et al. 2000) whereas an
upper limit for the secondary star temperature of 1700 K has been obtained
by  Ciardi et al. (1998). These spectral signatures along with the
extremely small difference between superhump and orbital period (Patterson
2001),  strongly suggests that the star is placed beyond the orbital period
minimum. The discovery of emission lines of the secondary during the 2001
superoutburst placed an upper limit on the secondary mass of 0.11
M$\sun$, but did not rule out the hypothesis of a
non degenerate mass donor (Steeghs et al.\,2001).

Our infrared spectroscopy, obtained outside eclipse,  shows a rich emission line spectrum, as
previously found by  Dhillon et al. (2000), Littlefair et al. (2000)
and by Mason et al. (2000). We identify Paschen $\beta$ double emission
with a half-peak separation of 610 km/s and FWZI
of 3900 km/s.  The Bracket series is also present as double emission, with
the violet peak larger than the red one, except in the Br$\gamma$ line
where the asymmetry reverses. Our $J$ magnitude, 14.25, compares well with the
$J$ magnitude 14.2 $\pm$ 0.2 given by Ciardi et al. (1998). Our $H$
magnitude, 14.07, is within the uncertainties of the 13.8 $\pm$ 0.2
value given by Ciardi et al. (1998). Our $K$ magnitude, 13.78, is lower
than the 13.3 $\pm$ 0.2 value reported by Ciardi et al. (1998). 
This is the only case where a significant difference exists
between one of our spectrophotometric magnitudes and previously published
values.

A simple power law provides the best fit for this object (Table 4).
By looking at the scaled template line strength in the K band
we estimate that the late type component contributes with less than 25 \% to the
flux in the $K$ band for the range M9.5-L7. By assigning such upper limit
to the companion star emission and using the astrometric parallax
of 45 pc (J. Thorstensen 2001, private communication), we estimate $M_{K}$
$>$ 12 for the secondary. This figure corresponds to a secondary with
a mass lower than or equal to the hydrogen burning minimum mass (0.07
M$\sun$), for ages between 0.1 and 10 Gyr, accordingly to the models of
very-low-mass stars and brown dwarfs with dusty atmospheres by Chabrier et
al. (2000).

\subsection{CU Vel} This object is a rather typical dwarf nova
with orbital period of  0.0785d and moderate outburst credentials (e.g.
Mennickent \& Diaz 1996). The Lyman alpha  absorption has been modeled as
photospheric emission from a white dwarf with effective temperature of
18500 K (G\"{a}nsicke \& Koester 1999).   Fig.\,5 shows that the best SED
model for \hbox{CU Vel} is reached with a M5 type spectrum and almost
zero contribution from a power law continuum. As in the
case of \hbox{RZ Leo}, M3 and M7 subtype templates failed to match the
shape of the spectrum, giving, in both cases,  $\chi^{2}$ about three times
larger than for the M5 template. We do not see the absorption lines of the
secondary in the J-band spectrum probably due to our poor S/N in this
range. The M5 spectral type is consistent with the 0.15 M$_{\odot}$
secondary derived from the dynamical analysis of the H$\alpha$ emission
line by Mennickent \& Diaz (1996). Following the same procedure as for
\hbox{RZ Leo}, we find  a distance of 150 $\pm$ 50  pc.

\subsection{HV Vir}
This is the faintest target in our sample, with magnitudes
18.7 and 17.0 in J and K respectively.  Our 30 min integration
time resulted in a very noisy spectrum. However, the spectrum
seems to indicate line emission in
Pa$\beta$, and a rather flat continuum.
The object has been linked to the sparsely populated
group of WZ Sge stars (e.g.\  Leibowitz et al. 1994).
Our SED algorithm reveals that a
L5 a late type template contributing with about 39\% to the total light
at 2.17$\mu$m gives a slightly better fit than a simple power
law. With the addition of a L5 template we get a
$\chi^{2}$ parameter slightly lower than for
L3 and L7 templates.
The $\chi^{2}$ is larger for power laws combined with other
spectral types.
The differences found in $\chi{2}$ are so small that
we should quote the possible
detection of a L5 type spectral contribution in \hbox{HV Vir}
with extreme caution.
Using the effective temperatures of L type dwarfs by Leggett et al. (2001),
one find $T_{eff}$ = 1550 $\pm$ 200 $K$ for the secondary of \hbox{HV
Vir}, and a distance of 175 $\pm$ 15 pc.

\subsection{1RXSJ105010.3}
This ROSAT source was recently identified as a cataclysmic variable star,
with an optical spectrum closely resembling WZ Sge, and likely
containing a sub-stellar secondary (Mennickent et al. 2001). The infrared
spectrum is very similar to that of WZ Sge too, both in the shape
of the continuum as well as in the rich emission line spectrum
and line strengths. No traces of the secondary spectral features
were found.

As in the case of \hbox{WZ Sge}, our SED models reveal
than a simple power law provides the
best fit.
This fact may suggest that, in spite of the low mass transfer rate,
the accretion disk is more
luminous than the secondary even in this spectral region.

\section{Discussion}

In Fig.\,6 we show the J-H:H-K color diagram for
short-period dwarf novae, along with data for
M-L type dwarfs. Colors for pure power law
energy distribution are also included for comparison.
We observe that the CV colors are distributed
in a rather ordered way, following a well-defined
track. Longer period systems (\hbox{RZ Leo}, \hbox{CU Vel})
show redder colors, above the region of the stellar photospheres,
whereas systems around the orbital period minimum are bluer,
and best represented by a simple power law continuum
with indexes between -2.8 and -1.8.

The most prominent emission lines are Paschen and Bracket lines. Helium lines
were not observed in our (small) sample. 
Among the observed spectra, \hbox{WZ Sge}, \hbox{1RXSJ105010.3},
and \hbox{VY Aqr} show the strongest emission
lines. These three systems are supposed to sustain very low mass transfer rates
during
quiescence, as may be inferred from their long recurrence times
and large amplitude outbursts.

In this paragraph we compare our results on the secondary star
of short orbital period CVs with the evolutionary predictions of
the population synthesis code
by Howell et al. (2001). At this point we emphasize the caveat already made
concerning the reliability of secondary temperature estimates from
the observed emission of a particular spectral type.
The Fig.\,7 shows a comparison of the
$T_{eff}-P_{orb}$  CV evolution tracks near the
orbital period minimum with our data and some additional
data taken from the literature. From this figure we conclude the
following: 1- \hbox{HV Vir}, \hbox{WZ Sge}, \hbox{EF Eri}, \hbox{WX
Cet}, \hbox{LL And} and \hbox{SW UMa} seem to be  post-orbital period
minimum systems. 2- It is difficult to reconciliate the positions
\hbox{VY Aqr} in the diagram with
the code's predictions.
3- In the same context, \hbox{RZ Leo} and \hbox{CU Vel} should be evolving
toward the orbital period minimum.

The results from the application of the spectral fitting procedure
suggests that the infrared continuum shape
in short-period cataclysmic variables may be
an useful indicator of the companion spectral type.
This point is
specially important if we consider that, due to the
limitation imposed by the spectrum S/N in such faint systems,
it is not always
possible to detect the secondary star individual lines,
but do the continuum shape. Also, the method have the
advantage of avoiding the uncertainties associated
with non-simultaneous multi-wavelength observations,
although their predictive power might be not so good
as the ideal case of modeling of simultaneous multi-wave-band observations
(e.g. Mason 2001).
In the future, we plan to observe a larger sample of CVs
which are candidates
for harboring brown dwarf like secondaries. In this regard,
the application of the method
to below-the-gap polars at deep quiescence turns to be especially
attractive (Howell \& Ciardi, 2001).
In these cases, the accretion of matter by the white dwarf virtually stops,
resulting in the easier detection of the secondary star at infrared
wavelengths.
In addition, a more realistic description of the disk emission in the IR
will be incorporated to the analysis.

\section{Conclusions}

\begin{itemize}
\item We have found some evidence for a M5 type spectral distribution in
\hbox{RZ Leo} and \hbox{CU Vel}.
This may suggest that these systems are in the upper branch of the
CV evolution track, i.e.\ prior to the orbital period minimum;

\item We find marginal evidence for a L5 type secondary star in
\hbox{HV Vir}. A secondary star with spectral type
M9.5 may be identified in \hbox{VY Aqr}. These systems have probably
passed beyond the orbital period minimum. However, the
position of \hbox{VY Aqr} in the $T_{2}$-$P_{orb}$ diagram
conflicts with the results of the population synthesis code by
Howell et al. (2001).
\item For some objects, namely \hbox{V485 Cen},
\hbox{WZ Sge} and \hbox{1RXSJ105010.3},
we found no significant improvement of
the spectral fitting by adding a stellar atmosphere template.
This may indicate that
even for such low luminosity systems the accretion disk spectrum
dominates the flux in the IR.
\end{itemize}

\section*{Acknowledgments}
We thanks the referee, Dr. Stuart Littlefair,
for useful  comments about a first 
version of this paper. We also thanks Sandy Leggett, 
who provided the IR digital
spectra of low mass
red objects used in our SED models.
We thanks Steve Howell and Elena Mason for providing
digitalized versions of the
CV evolution tracks shown in Fig.\,7.
Thanks to Claus Tappert
and Linda Schmidtobreick for stimulating discussions on a
preliminary version of this paper.
This work was supported by Grant
Fondecyt 1000324, DI UdeC 202.011.030-1.0,  CNPq 301029 and
FAPESP 99-06261.
We also acknowledge
support by grant Fundacion Andes C-13600/5.

\clearpage

\newpage

{\bf Figure Captions}

Fig.1 Combined J,H,K spectrum for selected short orbital period
  CVs. Fluxes are normalized by the factors shown in Table
2.

Fig.2 Comparison of the normalized spectrum of VY Aqr 
with the late-type templates.

Fig.3 The infrared spectrum of VY Aqr and the best composite SED fit
(thick line). The individual template spectrum and power law continuum are shown.
The spectra are scaled accordingly to the factor listed in Table 2.
A zoom around the \hbox{K\,{\sc i}} doublets is also shown in the upper left
graph. Main emission lines and absorption bands were excluded from the
fit.

Fig.4 The infrared spectrum of RZ Leo
and the best SED fit (thick line), corresponding to a M5 type
template.
The spectra are scaled   to the flux at $\lambda$ = 1.7 $\mu$m.

Fig.5 The infrared spectrum of CU Vel
and the SED best fit (thick line),
 corresponding to a M5 type
template.
The spectra are scaled to the flux at $\lambda$ = 1.7 $\mu$m.

Fig.6 The J-H:H-K diagram for short-period CVs
(solid triangles), M1-M6.5 dwarfs (open triangles,
from Leggett et al. 2000) and late-M and L-dwarfs (open circles,
from Leggett et al. 2001 and Reid et al. 2001).
Data points corresponding to pure power
law models are labeled accordingly to the power index
and they are interpolated by a dotted straight line.
CV data are from Table 2.

Fig.7 Cataclysmic variables close to the orbital period minimum.
Observations are compared with results of the CV population
synthesis code by Howell et al. (2001). We have used
spectral type - temperature calibrations
based on data of M-L dwarfs by Leggett et.
al.\,(2000, 2001).
Data for LL And and EF Eri
are from Howell \& Ciardi (2001),
for WX Cet, EF Peg and SW UMa
from Mason (2001) and
for WZ Sge (a temperature upper limit) from Ciardi et al. (1998).

\begin{table}
\caption[]{Journal of Observations. The total integration time, in seconds,
is given for every spectral band.}   \begin{center}
\begin{tabular}{lrccc} \hline \multicolumn{1}{c}{Object}&
\multicolumn{1}{c}{UT-Date} & \multicolumn{1}{c}{J} &
\multicolumn{1}{c}{H} & \multicolumn{1}{c}{K} \\  \hline VY Aqr
&23/05/00  &300 & 600 &1200\\ V485 Cen &21/05/00 &600 & &600 \\
RZ Leo   &21/05/00 &840 & 840 &840 \\ WZ Sge
 &20/05/00 & 240 & 480 &720   \\ CU Vel   &22/05/00 &120 & 240 &480 \\ HV
Vir   &19/05/00 &1800 &     &1800 \\
1RXSJ105010.3  &21/05/00 &240 &480 &720 \\   \hline \end{tabular}
\end{center} \end{table}

\begin{table}
\caption[]{Synthetic magnitudes and normalization factors
used in Fig.\,1, in units of
10$^{-17}$ erg cm$^{-2}$ s$^{-1}$ \AA$^{-1}$.}
\begin{center} \begin{tabular}{lrccc} 
\hline \multicolumn{1}{c}{Object}&
\multicolumn{1}{c}{J} &
\multicolumn{1}{c}{H} &
\multicolumn{1}{c}{K} &
\multicolumn{1}{c}{factor} \\ \hline

VY Aqr   &15.17& 14.66&14.14&29.7 \\
V485 Cen &16.47&     &15.30&9.16 \\
RZ Leo   &16.57& 15.68&15.45&8.89 \\
WZ Sge   &14.25& 14.07&13.78&74.0\\
CU Vel   &14.74& 13.94&13.51&46.6\\
HV Vir   &18.68&     &16.98 &1.06\\
1RXSJ105010.3 &17.07& 16.77&16.07&5.20\\

\hline  \end{tabular}
\end{center} \end{table}

\begin{table}
\caption[]{Equivalent width ($W(\lambda)$ in \AA) and peak flux (
$F(\lambda)$ in units of 10$^{-16}$
erg s$^{-1}$ cm$^{-2}$ \AA$^{-1}$) of the main emission lines.}
\begin{center} \begin{tabular}{lcccc} 
\hline \multicolumn{1}{c}{Object} &
\multicolumn{1}{c}{W(Pa$_{\beta})$} &
\multicolumn{1}{c}{F(Pa$_{\beta})$} &
\multicolumn{1}{c}{W(Br$_{\gamma})$} &
\multicolumn{1}{c}{F(Br$_{\gamma})$} \\ \hline

VY Aqr &-52 &4.79 &-59 &1.45  \\
V485 Cen &-18 &1.01 & &  \\
RZ Leo &-34 &1.01 &-37 &0.37  \\
WZ Sge & -144 & 17.3 &-142 & 2.76 \\
CU Vel & -41 & 6.44 & -26 &2.30  \\
1RXSJ105010.3 &-107 &1.00 &-142 &0.32  \\

\hline \end{tabular}
\end{center} \end{table}

 \begin{table}
\caption[]{Summary of the best SED models. We give
the best template spectral type (Sp2), their relative
contribution at $\lambda$ 2.17 $\mu$m (in \%) and the
parameter $c$.
N/A means not applicable.}
\begin{center} \begin{tabular}{lccc} 
\hline \multicolumn{1}{c}{Object}&
\multicolumn{1}{c}{Sp2} &
\multicolumn{1}{c}{rel. cont.} &
\multicolumn{1}{c}{$c$} \\ \hline
 VY Aqr    &M9.5   &44   &-2.2  \\
V485 Cen  &N/A &0   &-1.7 \\
 RZ Leo   &M5 &100 &N/A  \\
WZ Sge   &N/A &0 &-2.7 \\
CU Vel   &M5 &100 &N/A   \\
  HV Vir         &L5  &39 &-0.9 \\
 1RXSJ105010.3 &N/A &0 &-2.1 \\
\hline \end{tabular}
\end{center} \end{table}

\bsp

\label{lastpage}


\begin{thebibliography}{}
\bibitem[]{}Allen 2000, Allen´s Astrophysical Quantities, Arthur N. Cox (ed), AIP Press, Springer.
\bibitem[Augusteijn(1994)]{1994A&A...292..481A} Augusteijn, T.\, 1994, A\&A,
292, 481
\bibitem[Augusteijn(1994)]{} Augusteijn, T.\, 1996, A\&A,
311, 889


\bibitem[]{}Carter, B.\,S., \& Meadows, V.\,S.\, 1995, MNRAS, 276, 734

\bibitem[]{}Chabrier, G., Baraffe, I., Allard, F., Hauschildt, P.\, 2000, ApJ 542, 464

 \bibitem[Ciardi, Howell, Hauschildt, \& Allard(1998)]{1998ApJ...504..450C}
Ciardi, D.\,R., Howell, S.\,B., Hauschildt, P.\,H., \& Allard, F.\ 1998,
ApJ, 504, 450
\bibitem[G\"{a}nsicke \& Koester(1999)]{1999A&A...346..151G} G{\"
a}nsicke, B.\,T., \& Koester, D.\, 1999, A\&A, 346, 151

\bibitem[]{}Dahn, C. et al. 2000, Giant Planets to Cool Stars, ASP Conf. Ser. C. Griffith
and M. Marley eds.

\bibitem[Dhillon et al.(2000)]{2000MNRAS.314..826D} Dhillon, V.\ S.,
Littlefair, S.\,P., Howell, S.\,B., Ciardi, D.\,R., Harrop-Allin, M.\,K.,
\& Marsh, T.\,R.\, 2000, MNRAS, 314, 826

\bibitem[]{} Dhillon, V.\,S., Marsh, T.\,R.\, 1995, MNRAS 275, 89
 
\bibitem[]{} Downes, R.\,A., Webbink, R.\,F., Shara, M.\,M., et al. 2001
astro-ph/0102302   (see also http://icarus.stsci.edu/~downes/cvcat/)
 
\bibitem[]{}Howell, S.\,B.\, 2001, PASJ, 53, 675

\bibitem[Howell, Nelson, \& Rappaport(2001)]{2001ApJ...550..897H} Howell,
S.\,B., Nelson, L.\,A., \& Rappaport, S.\, 2001, ApJ, 550, 897
 
\bibitem[]{}Howell, S.\,B., Ciardi, D.\,R.\, 2001, ApJ, 550, L57
\bibitem[]{}Ishioka, R.\,et al.\, 2001, PASJ, 53, 905

\bibitem[]{}King, A.\,R., Schenker, K.\, 2002 in proceedings of
The Physics of Cataclysmic Variables and Related Objects,
ASP Conference Series, in press.
\bibitem[]{}Leggett, S.\,K., Allard, F., Dahn, C., Hauschildt, P.\,H., Kerr, T.\,H., \& Ryner, J.\,
2000, ApJ, 535, 965.
\bibitem[]{}Leggett, S.\,K., Allard, F., Geballe, T.\,R., Hauschildt, P.\,H., \& Schweitzer, A.\, 2001, ApJ 548, 908.

\bibitem[]{} Leibowitz, E.\,M., Mendelson, H., Bruch, A., et al. 1994, ApJ, 421, 771.
\bibitem[Littlefair, Dhillon, Howell, \& Ciardi(2000)]{2000MNRAS.313..117L}
Littlefair, S.\,P., Dhillon, V.\ S., Howell, S.\,B., \& Ciardi, D.\,R.\,
2000, MNRAS, 313, 117


\bibitem[]{}Mason, E.\, 2001, PhD thesis, Department of Physics
\& Astronomy,
University of Wyoming

\bibitem[Mennickent \& Diaz(1996)]{1996A&A...309..147M} Mennickent, R.\,E.,
\& Diaz, M.\, 1996, A\&A, 309, 147

\bibitem[]{}Mennickent, R.\,E., Sterken, C.\,W. Gieren\,E. Unda\,
1999, A\&A, 352, 239

\bibitem[]{}Mennickent, R.\,E., Diaz, M., Skidmore, W., Sterken, C.\,
2001, A\&A, 376, 448.
\bibitem[Mennickent \& Tappert(2001)]{1996A&A...309..147M} Mennickent, R.\,E.,
\& Tappert, C.\, 2001, A\&A, 372, 563
 
\bibitem[]{}Munari, U., \& Zwitter, T.\, 1998, A\&AS, 128, 277
 
\bibitem[Osaki(1996)]{1996PASP..108...39O} Osaki, Y.\, 1996, PASP, 108, 39

\bibitem[]{}Patterson, J.\, 2001, PASP 113, 736



\bibitem[]{}Reid, I.\,N., Burgasser,
A.\,J., Cruz, K.\,L., et al. 2001, AJ 121, 1710

\bibitem[]{}Rousselot, P., 
Lidman, C., Cuby, J.-G., Moreels, G., \& Monnet, G.\, 2000, A\&A, 354, 1134.


\bibitem[Sproats, Howell, \& Mason(1996)]{1996MNRAS.282.1211S} Sproats, L.\,
N., Howell, S.\,B., \& Mason, K.\,O.\, 1996, MNRAS, 282, 1211

\bibitem[]{}Steeghs, D., Marsh, T., 
Knigge, C., Maxted, P.~F.~L., Kuulkers, E., \& Skidmore, W.\, 2001, ApJ, 562L, 145
 
\bibitem[Thorstensen \& Taylor(1997)]{1997PASP..109.1359T} Thorstensen, J.\,
R., \& Taylor, C.\, J.\, 1997, PASP, 109, 1359

 

\bibitem[]{}van Altena, W.\,F., Lee, J.\,T., \& Hoffleit, E.\,D.\, 1994, The General Catalogue of
Trigonometric Parallaxes (New Haven: Yale University Observatory)

\end{thebibliography}
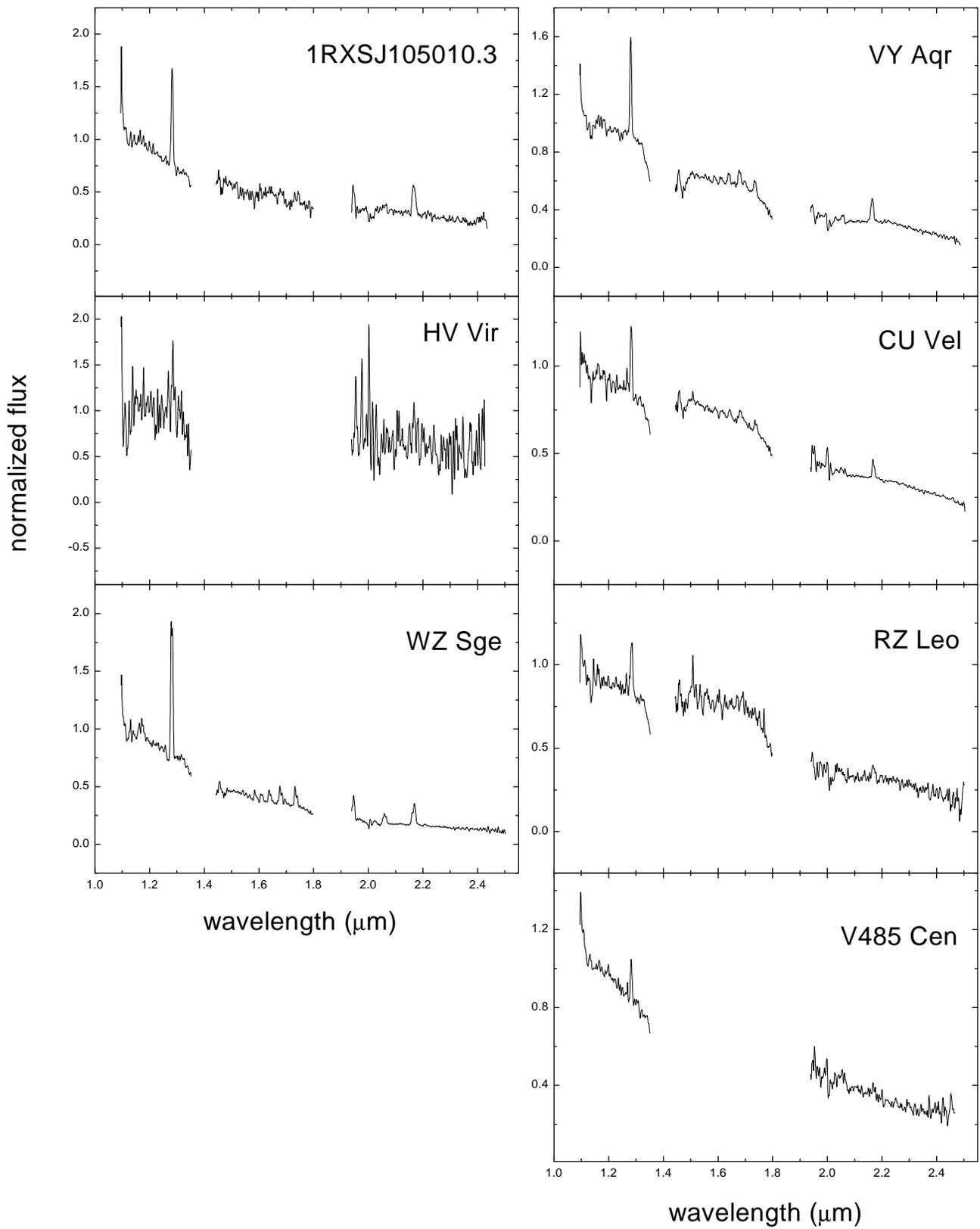
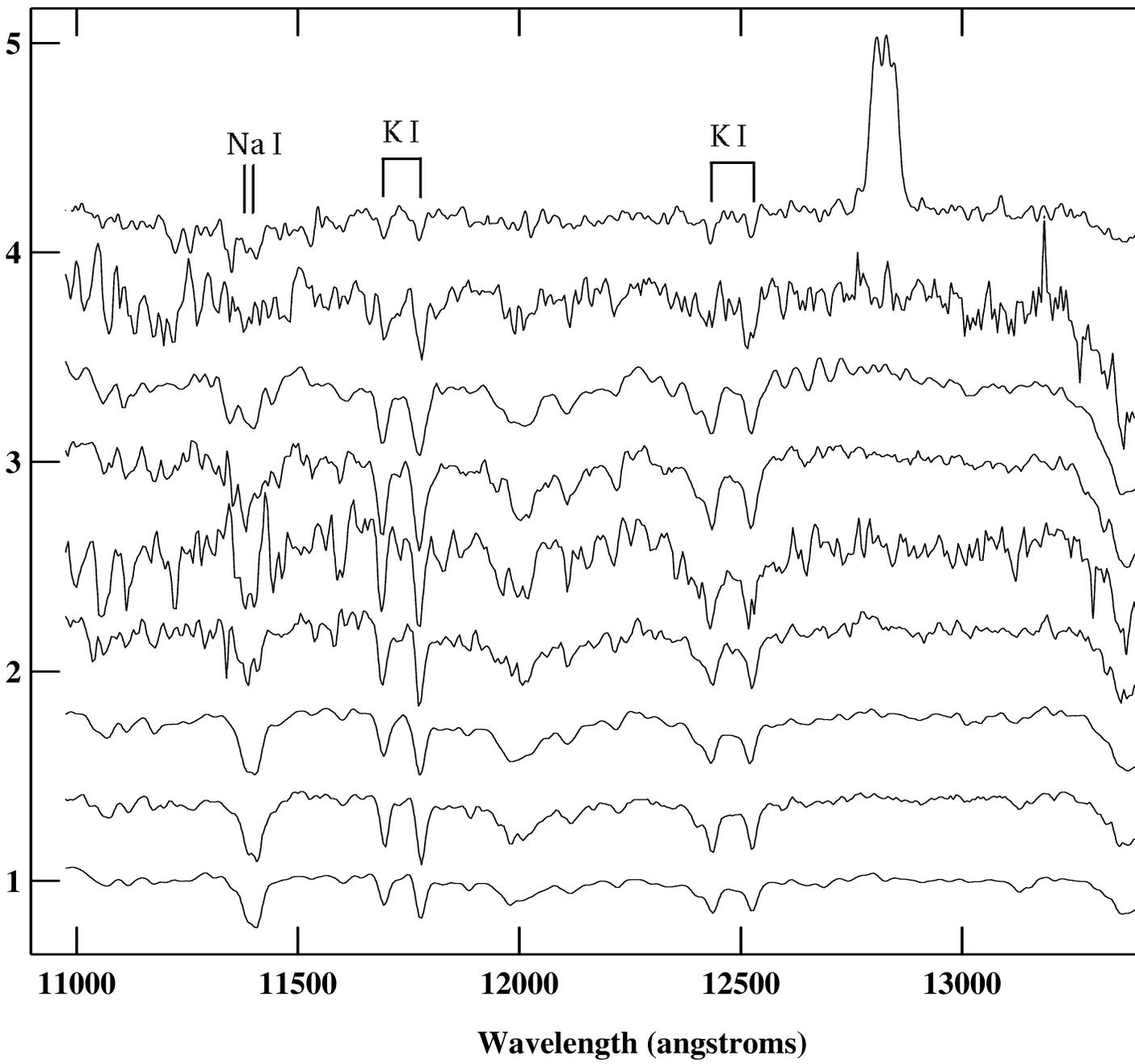
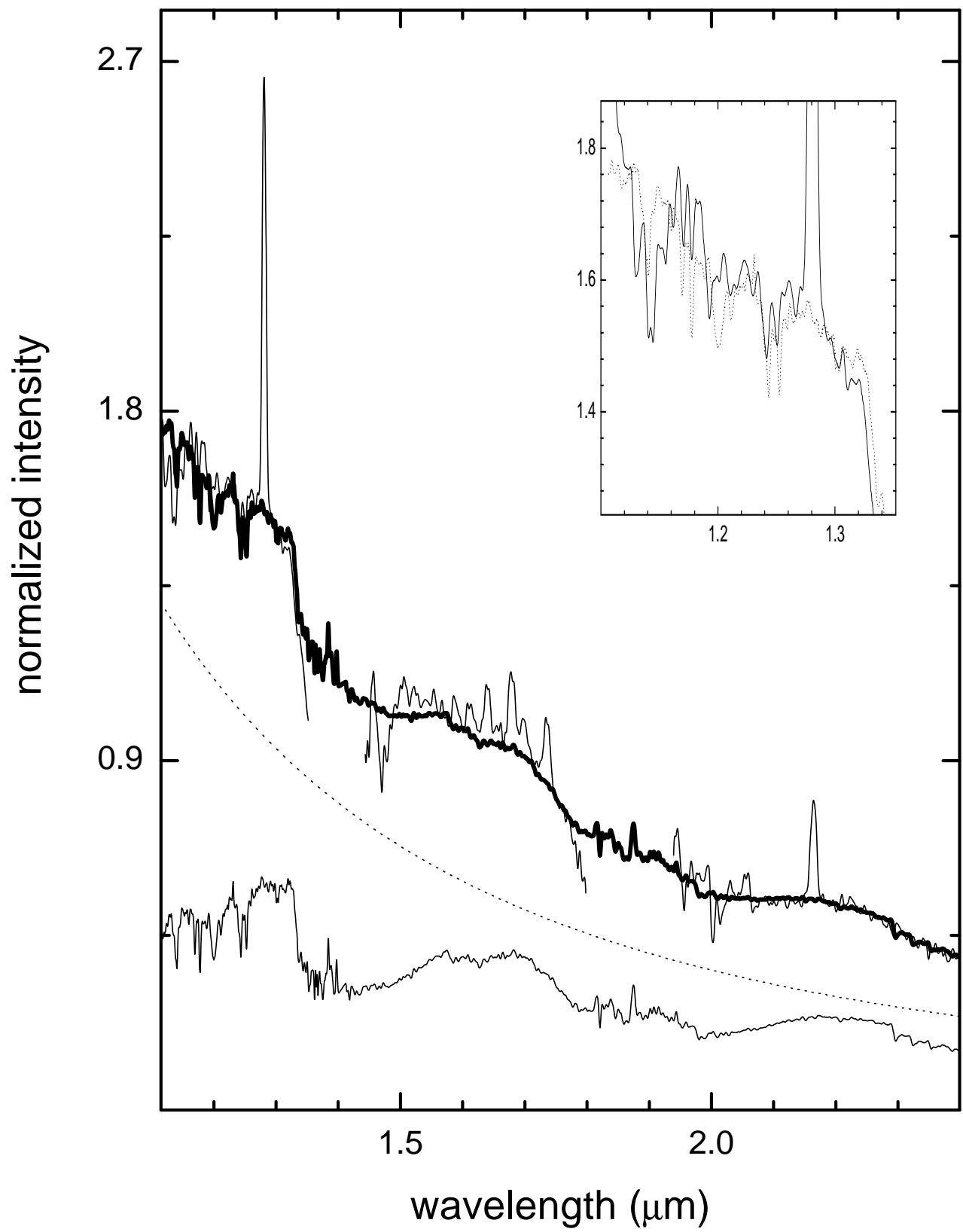
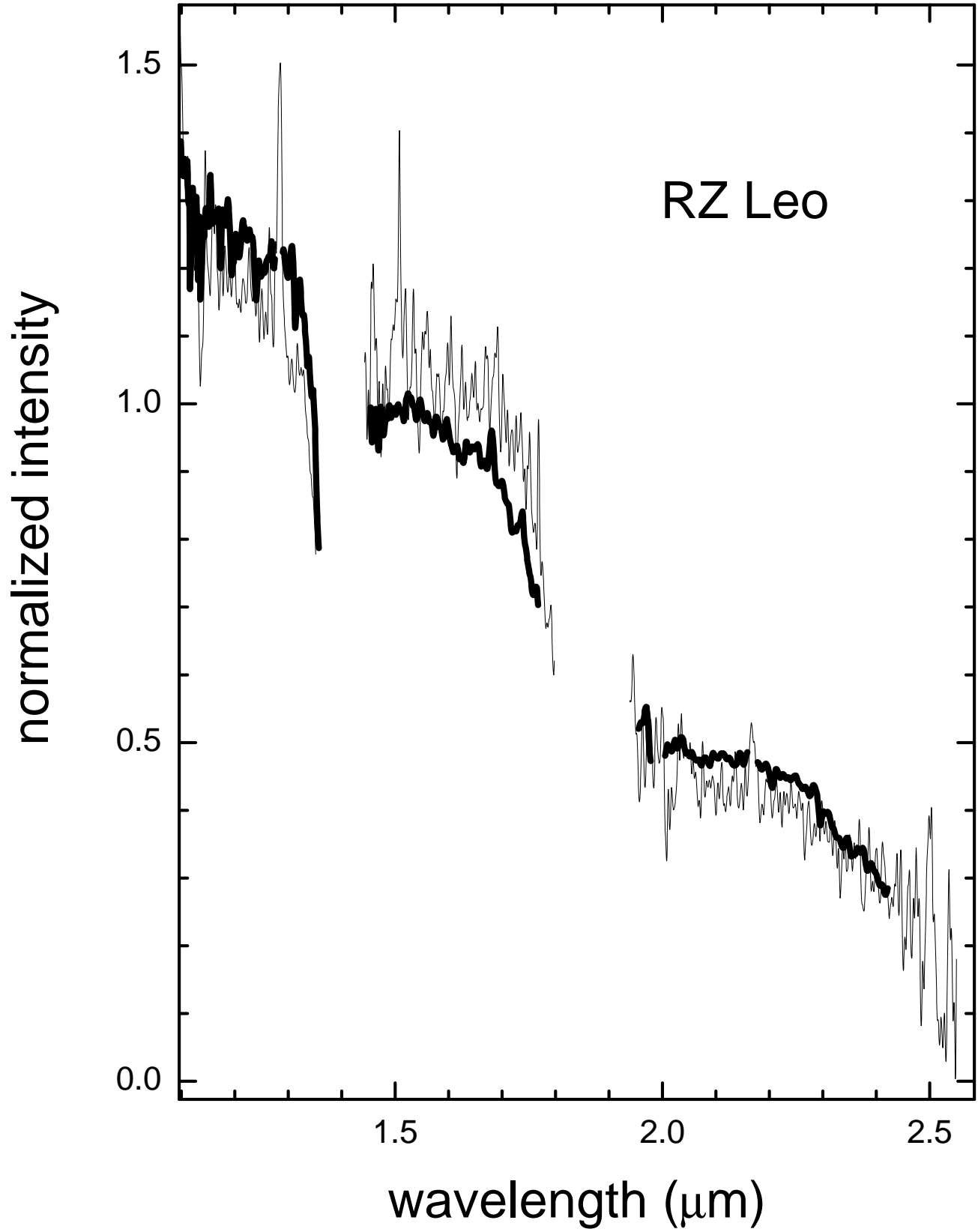
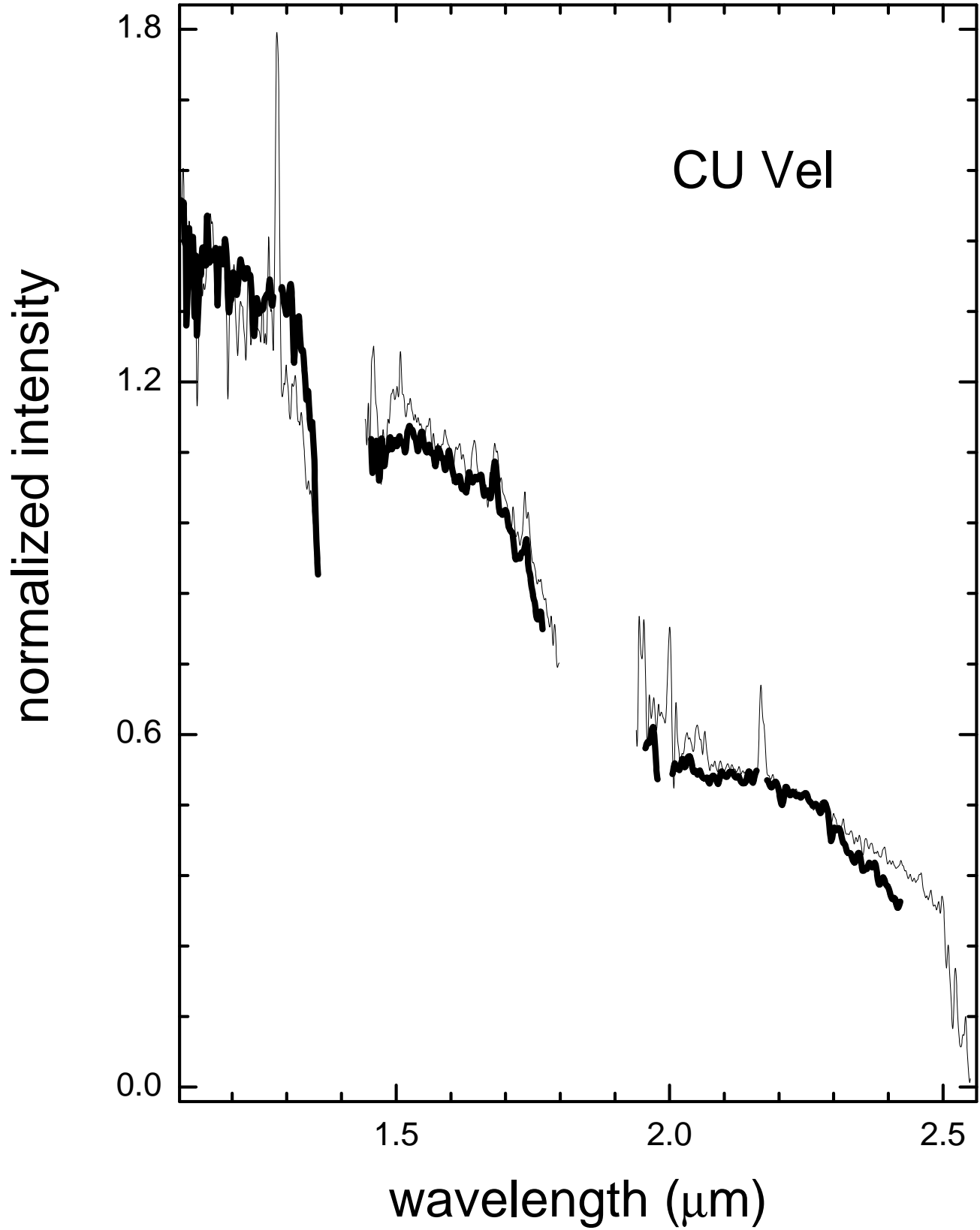
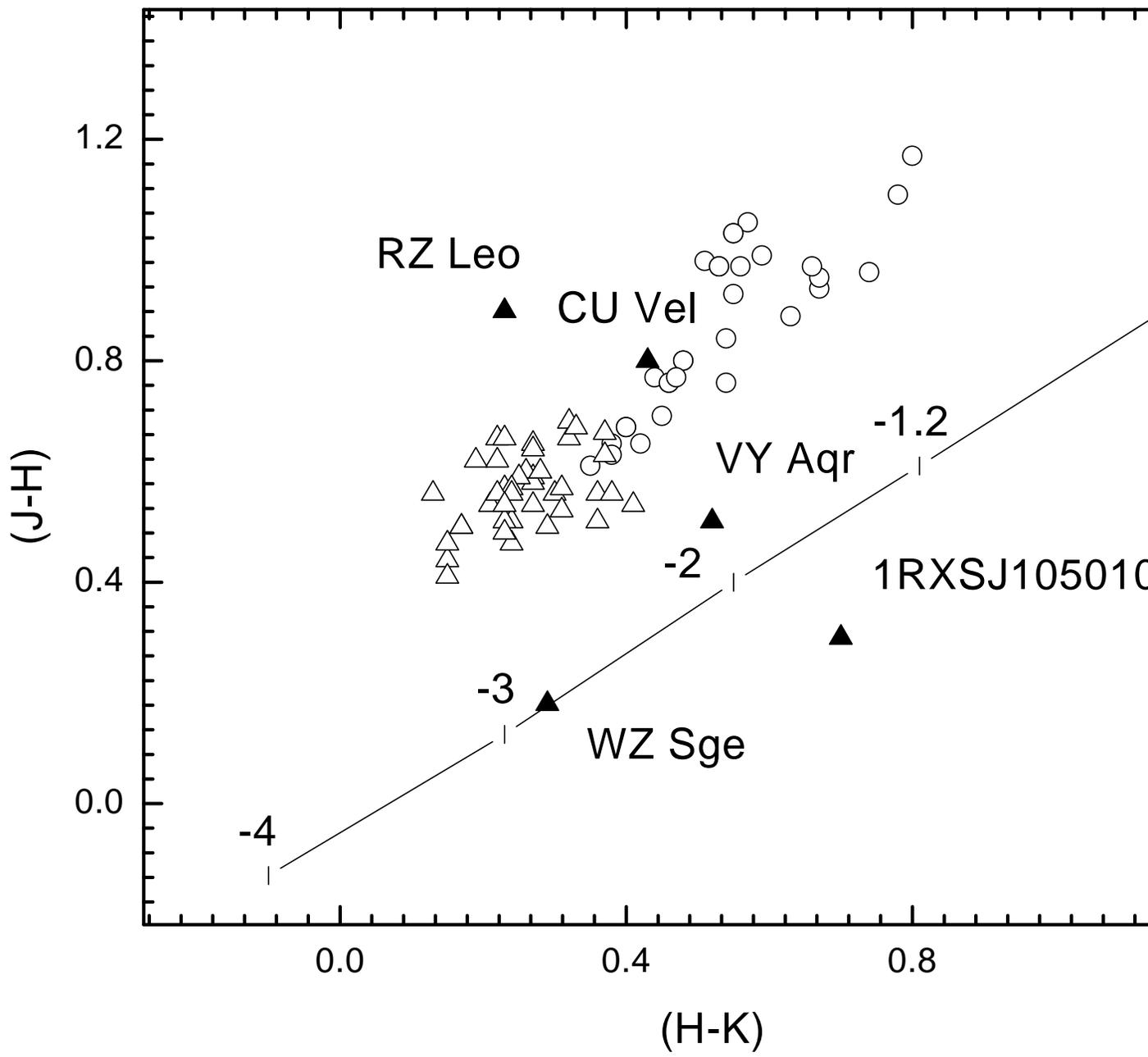
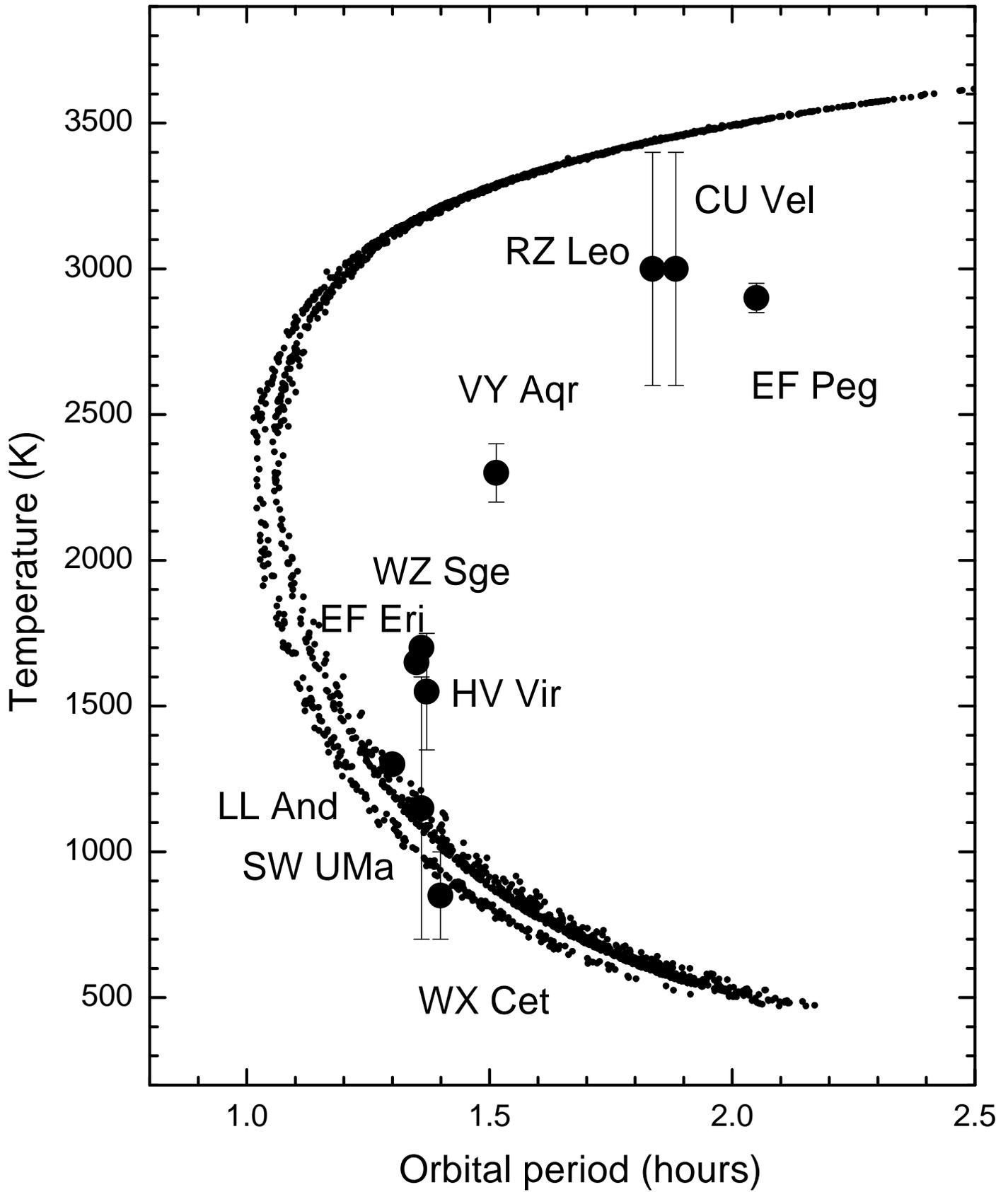
\end{document}